\documentclass[preprint,pra,aps]{revtex4}

\usepackage{epsfig}

\begin{document}

\author{Daniel Loss$^1$ and David P. DiVincenzo$^2$}

\title{Exact Born Approximation for the Spin-Boson Model}

\affiliation{\vspace*{1.2ex}$^1$
            \hspace*{0.5ex}{Department of Physics and Astronomy, University of Basel,
            Klingelbergstrasse 82, CH-4056 Basel, Switzerland}}
\affiliation{\vspace*{1.2ex}$^2$
            \hspace*{0.5ex}{IBM Research Division, T. J. Watson Research Center,
P.O. Box 218, Yorktown Heights, NY 10598, USA}}

\date{\today}

\begin{abstract}

Within the lowest-order Born approximation, we present an exact
calculation of the time dynamics of the spin-boson model in the
Ohmic regime.  We observe non-Markovian effects at zero
temperature that scale with the system-bath coupling strength and
cause qualitative changes in the evolution of coherence at
intermediate times of order of the oscillation period.  These
changes could significantly affect the performance of these
systems as qubits. In the biased case, we find a prompt loss of
coherence at these intermediate times, whose decay rate is set by
$\sqrt{\alpha}$, where $\alpha$ is the coupling strength to the
environment.  These calculations indicate precision experimental
tests that could confirm or refute the validity of the spin-boson
model in a variety of systems.


\end{abstract}

\maketitle

Novel solid state devices that can control spin degrees of freedom
of individual electrons\cite{LK,DL97}, or discrete quantum states
in superconducting circuits\cite{Naka,delft,vion,RMPkarl}, show
promise in realizing the ideal of the completely controllable
two-state quantum system, weakly coupled to its environment, that
is the essential starting point for qubit operation in quantum
computation.  From a fundamental point of view, these experimental
successes also bring us close to embodying the ideal test of
quantum coherence as envisioned by Leggett many years
ago\cite{Leggett}, in which a simple quantum system is placed in a
known initial state, is allowed to evolve for a definite time $t$
under the action of its own Hamiltonian and under the influence of
decoherence from the environment, and is then measured.

Recent experiments, starting with \cite{Naka}, show that this
ideal test can be implemented in practice.  The decay of quantum
oscillations due to environmental decoherence is
now\cite{Naka,delft,vion,RMPkarl} sufficiently weak that some tens
of coherent oscillations can be observed. If quantum computation
is to become a reality, it is believed\cite{DL} that these systems
will eventually need to achieve even lower levels of decoherence,
such that thousands or tens of thousands of coherent oscillations
could be observed.  This prospect of producing experiments with
ultra-long coherence times in quantum two state systems offers a
new challenge for theoretical modelling of decoherence.  Despite
the many years of work\cite{weiss,Haenggi} following on Leggett's
initial proposals, there has never been a full, systematic
analysis of the most popular description of these systems, the
spin-boson model, in the limit of very weak coupling to the
environment.

In this Letter we provide an exact analysis of the weak coupling
limit of the spin boson model for the Ohmic heat bath, and in the
low temperature limit.  In this limit the Born approximation (to
the self energy) should become essentially exact, and we make no
other approximations in our solutions --- in particular, no Markov
approximation is made. As other workers have recently
emphasized\cite{braun,privman}, understanding the details of the
short-time dynamics of this model is especially crucial for the
operation of these systems as qubits.

We find important, new, non-Markovian effects in this regime.  At
lowest order in the Born expansion of the self energy
superoperator, the time dynamics of the model rigorously separates
into a sum of strictly exponential pieces (the usual ``$T_1$" and
``$T_2$" decays of the Bloch-Redfield model) plus two distinct
non-exponential pieces that arise, technically speaking, from two
different kinds of branch cuts in the Laplace transform of the
solution of the generalized master equation that we obtain.

These two contributions both have power-law forms at long times,
$t>T_1,T_2$, and thus formally dominate the exponentially-decaying
parts.  But more interesting is that they both give new structure
to the time evolution at intermediate times $t$, $1/\omega_c <t<
T_1,T_2$; this structure typically occurs for $t$ on the order of
the oscillation period. (Here, $\omega_c$ is a high frequency
cut-off of the bath modes, defining the very short time regime,
$t<1/\omega_c$, which is of no interest here.)
 We can
explain our results in the language of the double-well potential,
where the two quantum states are ``left" and ``right" ($L$/$R$),
the $t=0$ state is pure $L$, and the system oscillates in time via
tunneling from $L$ to $R$.  The first branch-cut contribution is
most important in the unbiased case ($L$ and $R$ energies
degenerate) and it causes the system, starting immediately in the
first quantum oscillation, to spend more time in the $R$ well,
that is, the {\em opposite} well from the one the system is in
initially.  The second branch-cut contribution, present when the
system is biased, adds to the amplitude of the coherent
oscillation, but dies out after an intermediate time which scales
like the inverse square root of the interaction strength $\alpha$
with the bath.  This {\em prompt loss of coherence}, whose
amplitude is proportional to $\alpha$, changes qualitatively the
picture of the initial decay of coherence that is so important for
discussions of fault-tolerant quantum computation\cite{Preskill}.

We are interested in studying the time dependence of the system
density matrix $\rho_S(t)={\rm Tr}_B \rho(t)$ with a
time-independent system Hamiltonian, and in the presence of a
fixed coupling to an environment.  An exact equation for $\rho_S$
-- the generalized master equation (GME) -- is\cite{FS}
\begin{eqnarray}
\dot\rho_S(t)&=&-iL_S\rho_S(t)-i\int_0^tdt'\Sigma_S(t-t')\rho_S(t'),
\label{p1_3e1_11_1}\\
\Sigma_S(t)&=&-i{\rm Tr}_BL_{SB}e^{-iQLt}L_{SB}\rho_B.
\label{p1_3e1_11_2}
\end{eqnarray}
Here the kernel $\Sigma_S(t)$ is the self energy superoperator,
the system-bath Hamiltonian is written $H=H_S+H_{SB}+H_B$,
the Liouvillian superoperator is defined by $L_x\rho=[H_x,\rho]$,
$\rho_B=e^{-\beta H_B}/Z$, $\beta=1/k_BT$, $T$ is the temperature,
and $Q$ is the projection superoperator $Q=1-\rho_B{\rm Tr}_B$.
Eq. (\ref{p1_3e1_11_1}) is written for the case ${\rm
Tr}_BH_{SB}\rho_B=0$, and the total initial state is taken to be
of the form $\rho(0)=\rho_S(0)\otimes\rho_B$.  Since we are
interested in the case of weak coupling to the bath, we will
consider a systematic expansion in powers of this coupling $L_{SB}$ in the
self-energy operator $\Sigma_S(t)$. Retention of only the lowest order
term in this expansion, giving the Born approximation, is
obtained\cite{CL} by the replacement
$e^{-iQLt}\rightarrow e^{-iQ(L_S+L_B)t}$
in Eq. (\ref{p1_3e1_11_2}).

We now proceed to solve the GME {\em with no further
approximations.}  This distinguishes our work from previous
efforts, in which various other approximations (secular, rotating
wave, Markov, ``non-interacting blips", short time) are made (see,
e.g., \cite{Leggett,Haenggi,weiss,braun,privman}). We will find
that, in particular, avoidance of the Markov approximation endows
the solution with qualitatively new features.

We obtain our solution for the special case of a two-dimensional
system Hilbert space, and a system-bath coupling of a simple
bilinear form, $H_{SB}=S\otimes X$
($S$ ($X$) is an operator in the system (bath) space).  In this
case the GME (\ref{p1_3e1_11_1}) in Born approximation
can be rewritten in
an ordinary operator form:
\begin{eqnarray}
\langle\dot\sigma_\mu(t)\rangle&=&-i{\rm Tr}_S\sigma_
\mu[H_S,\rho_S(t)]-\int_0^tdt'I_\mu(t,t')\, ,\label{p2_2e2_61}\\
I_\mu(t,t')&=&I_{\mu
0}(t')+\sum_{\nu=1}^3I_{\mu\nu}(t')\langle\sigma_\nu(t-t')\rangle\, ,\label{p2_2e2_62}\\
I_{\mu\nu}(t')&=&{\rm Re}~\{C(-t'){\rm
Tr}_S\sigma_\nu(-t')[\sigma_\mu,S]S(-t')\}\, .\label{p2_2e2_63}
\end{eqnarray}
Here $\sigma_{x,y,z}$ are the Pauli operators, $\sigma_0=I$,
$\langle x\rangle\equiv{\rm Tr}_S x\rho_S$, the bath
correlation function is $C(t)\equiv {\rm
Tr}_B[XX(t)\rho_B]=C'(t)+iC''(t)$, the time dependent operators
are in the interaction picture, i.e.,
$o(t)=e^{i(H_S+H_B)t}oe^{-i(H_S+H_B)t}$,
and $C'$ and $C''$ denote the real and imaginary parts of the bath
correlator.

Without loss of generality, we can take the system operators to be
of the form
\begin{equation}
H_S={\Delta\over 2}\sigma_x+{\epsilon\over 2}\sigma_z
\end{equation}
and
$S=\sigma_z$.
Then the GME can be written in an explicit form
$\langle\underline{\dot\sigma}(t)\rangle=
R*\langle\underline{\sigma}\rangle+\underline{k}$.
Here $\underline{\sigma}$ denotes the vector
$(\sigma_x,\sigma_y,\sigma_z)^T$, convolution is denoted
$A*B\equiv\int_0^tdt'A(t')B(t-t')$, and
\begin{eqnarray}
R(t)&=&\left(\begin{array}{ccc}-{E^2\over\Delta^2}\Gamma_1(t)&
-\epsilon\delta(t)+{E\over\Delta}K^+_y(t)&0\\
\epsilon\delta(t)-{E\over\Delta}K^+_y(t)&-\Gamma_y(t)&-\Delta\delta(t)\\
0&\Delta\delta(t)&0\end{array}\right),\label{p2_10e2_51}\\
\underline{k}(t)&=&\left(\begin{array}{ccc}-{E\over\Delta}k^-(t),&
-k_y^-(t),&0\end{array}\right)^T,\label{p2_10e2_492}
\end{eqnarray}
with $E=\sqrt{\epsilon^2+\Delta^2}$,
and kernels given by $\Gamma_1(t)={4\Delta^2\over
E^2}\cos(Et)C'(t)$,
$\Gamma_y(t)={4\Delta^2\over
E^2}(1+{\epsilon^2\over\Delta^2}\cos(Et))C'(t)$,
$K_y^+(t)={4\epsilon\Delta\over E^2}\sin(Et)C'(t)$,
$k^-(t)={4\Delta^2\over E^2}\int_0^tdt'\sin(Et')C''(t')$, and
$k_y^-(t)={4\epsilon\Delta\over
E^2}\int_0^tdt'(1-\cos(Et'))C''(t')$\cite{foot1}.

These equations can be solved in the Laplace domain.  Defining the
Laplace transform as $f(s)=\int_0^\infty e^{-st}f(t)dt$,
the solutions are, for the ``standard" initial conditions
$\langle\underline\sigma(t=0)\rangle=(0,0,z_0=1)^T$,
\begin{eqnarray}
\langle\sigma_x(s)\rangle&=&{1\over
s+{E^2\over\Delta^2}\Gamma_1(s)}\left(\left(\epsilon-{E\over\Delta}K_y^+(s)\right){N(s)\over
D(s)}-{E\over\Delta}k^-(s)\right),\label{p2_11e2_541}\nonumber\\
\langle\sigma_y(s)\rangle&=&-{N(s)\over D(s)}\,\,\,
,\label{p2_11e2_542}\,\,\,\,\,\,\,\,
\langle\sigma_z(s)\rangle=-{\Delta\over s}{N(s)\over
D(s)}+{z_0\over s}\,\,\, ,\label{p2_11e2_543}\\
N(s)&=&{E\over\Delta}\left(\epsilon-{E\over\Delta}K_y^+(s)\right)k^-(s)+\left({\Delta\over
s}z_0+k_y^-(s)\right)\left(s+{E^2\over\Delta^2}\Gamma_1(s)\right),\label{p2_11e2_544}\\
D(s)&=&\left(s+\Gamma_y(s)+{\Delta^2\over
s}\right)\left(s+{E^2\over\Delta^2}\Gamma_1(s)\right)+
\left(\epsilon-{E\over\Delta}K_y^+(s)\right)^2.\label{p2_11e2_545}
\end{eqnarray}
We can give more explicit solutions if we specialize to the
spin-boson model, for which $X=\sum_n c_n(b_n^\dagger+b_n)$ and
$H_B=\sum_n\omega_nb_n^\dagger b_n$.
If the spectral function $J(\omega)$ is defined as
$J(\omega)\equiv\sum_nc_n^2\delta(\omega-\omega_n)$, then
\begin{equation}
C(t)=\int_0^\infty d\omega
J(\omega)(\coth({\beta\omega/2})\cos(\omega t)+i\sin(\omega
t)).\label{p7_2e48}
\end{equation}
For most of the sequel we will consider the Ohmic case, for which
$J(\omega)={\alpha\over 2}\omega e^{-\omega/\omega_c}$ ($\omega_c$
is an ultraviolet cutoff).  In this case Eq. (\ref{p7_2e48})
becomes
\begin{equation}
C(t)=-{\alpha\over\beta^2}{\rm
Re}~\psi'\left({1-i\omega_ct\over\beta\omega_c}\right)+{\alpha\omega_c^2\over
2}{1-\omega_c^2t^2\over(1+\omega_c^2t^2)^2}+i\alpha\omega_c^2
{\omega_ct\over(1+\omega_c^2t^2)^2}~~,\label{DDV25Oct2002}
\end{equation}
where $\psi'$ is the derivative of the digamma function\cite{AS}.

For discussing the exact solution it is instructive to understand
the structure of the solution in a Markov approximation.  This
approximation is obtained by replacing all the kernels $\Gamma_1$,
$\Gamma_y$, $K_y^+$, $k^-$, and $k_y^-$ by their values at $s=0$.
In this case the solutions Eqs. (\ref{p2_11e2_543}) are rational
functions of $s$. Then, if the poles of these rational functions
are located at positions $s_k$ in the complex $s$ plane, with
residues $r_k/2\pi i$, then the inverse Laplace transform can be
written $\langle\sigma_\mu(t)\rangle=\sum_k r_k^\mu \exp(s_i t)$.
We indicate here that while the residues do depend
on the label $\mu=x,y,z$, the pole positions do not, as is
suggested by the form of Eqs. (\ref{p2_11e2_543}).

As is well known\cite{weiss}, there are four poles: $s_1=0$,
$s_2=-\Gamma_1^0$, and $s_{3,4}=-\Gamma_2^0\pm i{\tilde E}$.  The
first pole describes the long-time asymptote of the solution (stationary state), the
second the purely exponential, ``$T_1$"-type decay (relaxation), and the last
two (complex conjugate paired) describe an exponentially decaying
sinusoidal part, the ``$T_2$"-type decay of coherent oscillations.  The residues of these
poles are, to lowest order in $\alpha$,
$r_1^x=x_\infty=-(\Delta/E)~\tanh(\beta E/2)$, $r_1^y=y_\infty=0$,
$r_1^z=z_\infty=-(\epsilon/E)~\tanh(\beta E/2)$,
$r_2^x=\epsilon\Delta/E^2-x_\infty$, $r_2^y=0$,
$r_2^z=\epsilon^2/E^2-z_\infty$, $r_{3,4}^x=-\epsilon\Delta/2E^2$,
$r_{3,4}^y=-\Delta/2E$, and $r_{3,4}^z=\Delta^2/2E^2$.
The pole positions are, again to lowest order in $\alpha$, given
by $\Gamma_1^0=T_1^{-1}={\alpha\pi\Delta^2\over E}\coth(\beta
E/2)$,
$\Gamma_2^0=T_2^{-1}=\Gamma_1^0/2+{2\alpha\pi\epsilon^2\over
E^2}k_BT$,
and ${\tilde E}=E+\delta E$, $\delta E=\delta E^{\rm Lamb}+\delta
E^{\rm Stark}$\cite{GS}, $\delta E^{\rm Lamb}={\alpha\Delta^2\over
E} (C + \ln(E/\omega_c))$
and $\delta E^{\rm Stark}={\alpha\Delta^2\over E}({\rm
Re}~\psi(iE\beta/2\pi)-\ln(E\beta/2\pi))$, where we have dropped
terms of order $E/\omega_c$ and higher, C is the Euler constant,
and $\psi$ is the digamma function \cite{AS}. These expressions
are straightforwardly derivable, and agree with the
literature\cite{weiss}, except for the energy shift due to
vacuum fluctuations, $\delta E^{\rm
Lamb}$ (which contains in general $\ln(E/\omega_c)$ and not
$\ln(\Delta/\omega_c)$).
 We note
that this Markovian theory satisfies the expected fundamental
relation
$\Gamma_2^0=\Gamma_1^0/2+(2\epsilon^2/E^2)\int_{-\infty}^\infty
dt\langle X(t)X\rangle_B$ (Korringa relation)\cite{Abragam}; also,
to lowest order in $\alpha$, the asymptotic values of
$\langle\sigma_\mu(t\rightarrow\infty)\rangle$ go to the Boltzmann
equilibrium distribution of the system, e.g.,
$z_\infty=-(\epsilon/E)~ \tanh(\beta E/2)$, unlike, for example,
the popular ``non-interacting blip" approximation\cite{weiss}.

We now return to the exact solution, examining it in detail at
vanishing temperature $T=0$.  In this case the Laplace transform of $C$ in Eq.
(\ref{DDV25Oct2002}) is
\begin{eqnarray}
C_{T=0}(s)&=&{\alpha s/2}\left(-\cos\left({\tilde s}\right){\rm
Ci}\left({\tilde s}\right)- \sin\left({\tilde s}\right){\rm
si}\left({\tilde
s}\right)\right)\nonumber\\\label{DDVmathematicac1anycut}
&-&{i\alpha/2}\left(-\omega_c + s~\sin\left({\tilde s}\right){\rm
Ci}\left({\tilde s}\right) - s\cos\left({\tilde s}\right){\rm
si}\left({\tilde s}\right)\right)\, ,\label{DDVmathematicac1anycut2}
\end{eqnarray}
where ${\tilde s}=s/\omega_c$\cite{footnotewc}.
There is an important
feature of this correlation function that makes the Markov solution
qualitatively incomplete: while the sine integral si is analytic,
the cosine integral Ci(s) behaves like $\ln(s)$ for $s\rightarrow
0$\cite{AS}. This means that $C(s)$ is nonanalytic at $s=0$ --- it
has a branch point there.  Thus, the exact solutions
$\langle\sigma_\mu(s)\rangle$ have extra analytic structure not
present in the Markov approximation, and the real-time dynamics
$\langle\sigma_\mu(t)\rangle$ has qualitatively different features
in addition to the pole contributions we have just discussed.

The $s=0$ branch point in $C(s)$ leads the kernels $\Gamma_1(s)$,
$K_y^+(s)$, and $k^-(s)$ to have branch points at $s=\pm iE$; the
kernels $\Gamma_y(s)$ and $k_y^-(s)$ have three branch points, at
$s=0$ and $s=\pm iE$.  Thus, the solutions to the GME
$\langle\sigma_{x,y,z}(s)\rangle$ also have three branch points in
the complex plane.  We find by numerical study that the exact
solutions still have four poles as before, which, for small
$\alpha$, have nearly (but not exactly) the same pole positions
and residues as in the Markov approximation.

\begin{figure}[htb]
\epsfxsize=15cm \epsffile{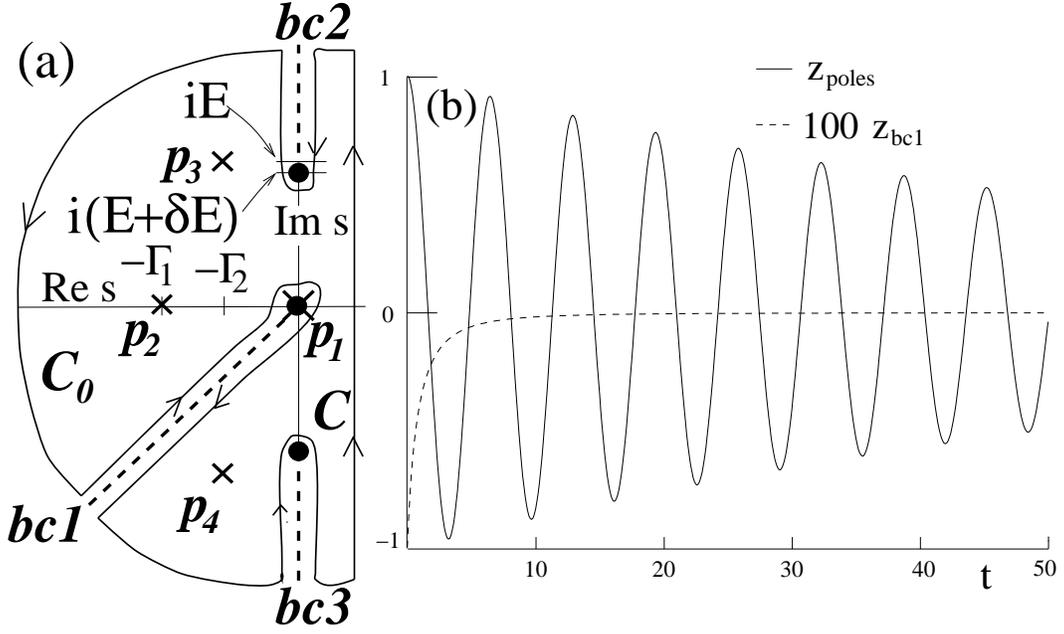} \caption{(a) Structure of
the solutions $\langle\sigma_\mu(s)\rangle$ in the complex $s$
plane.  The four poles $p_1$, $p_2$, $p_3$, and $p_4$, are
indicated by crosses; the three branch points at $s=0,\pm iE$ are
indicated by solid circles, and the three branch cuts chosen, bc1,
bc2, and bc3, are indicated by dashed lines.  The inverse Laplace
transform requires an integration along the contour ${\bf C}$
parallel to the imaginary axis.  This integral may be evaluated by
closing with a contour in the left half plane (${\bf C_0}$, the
Bromwich contour), which lies at infinity except for looping back
around each of the branch cuts. (b) $z_{poles}(t)$ and
$z_{bc1}(t)$ for the unbiased case, $\epsilon=0$, $\Delta=1$,
$\omega_c=30$, $T=0$, and $\alpha=0.01$.  $t$ is in units of $1/E$
(i.e., $E=1$).} \label{fig1}
\end{figure}

Thus, the structure of the solutions in the complex $s$ plane is
as shown in Fig. \ref{fig1}a.  The locations of the branch cuts are
chosen for computational convenience, as discussed shortly. Given
this branch-cut structure, the inverse Laplace transform (the
Bromwich integral) is evaluated by closing the contour as shown.
Thus, the exact inverse Laplace transform can be expressed as ($t>0$)
\begin{eqnarray}
\langle\sigma_\mu(t)\rangle&=&{1\over 2\pi i}\int_{\cal
C}dse^{st}\langle\sigma_\mu(s)\rangle={1\over 2\pi i}\oint_{{\cal
C}_o}dse^{st}\langle\sigma_\mu(s)\rangle\nonumber\\
&&-{1\over 2\pi i}\sum_{k=1}^3 q_k\int_{p_k}^{\infty}dxe^{q_kxt}
(\langle\sigma_\mu(q_kx+\eta_k)\rangle-
\langle\sigma_\mu(q_kx-\eta_k)\rangle). \label{DDV1_6_03}
\end{eqnarray}
Here $q_k=e^{i\theta_k}$ and $\eta_k=\eta e^{i(\theta_k-\pi/2)}$,
with $\eta$ an infinitesimal positive real number.  That is,
$\eta_k$ is an infinitesimal displacement perpendicular to the
direction of branch cut $k$.  For the cut choices we have made,
$\theta_1=5\pi/4$, $\theta_2=\pi/2$, $\theta_3=3\pi/2$, $p_1=0$,
and $p_2=p_3=E$. The closed-contour integral in the expression can
be written as a sum over the four poles, and so gives complex
exponential contributions to the solution as in the Markovian
case.  The extra terms, the sum over the three branch cuts, are
new and give qualitatively different features.  The contributions
of the second and third branch cuts are complex conjugates of each
other, so we will discuss them together as the ``branch cut 2"
(``bc2" for short) contribution.  The first cut will be discussed
as the ``bc1" contribution.

The contribution of these cuts to the solution is independent of
the detailed positioning of the branch cuts, so long as they are
not moved across a pole; the choice of the direction of bc1 is a
computational convenience --- the apparently most natural choice
of this cut direction, along the negative real axis, passes it
essentially on top of the $\Gamma_1$ pole, making the evaluation
of the branch-cut integral numerically inconvenient. As a check,
we find that the results we discuss now are indeed independent of
the cut direction.

We have done a detailed study of these branch-cut contributions
for $\langle\sigma_z(t)\rangle\equiv z(t)$.  For the unbiased
spin-boson case, $\epsilon=0$, an essentially analytic calculation
can be done for all contributions.  In this case there is no bc2
contribution, $z_{bc2}=0$.  The bc1 contribution can be obtained
analytically to leading order in $\alpha$:
$z(t)=z_{poles}(t)+z_{bc1}(t)$,
\begin{equation}
z_{bc1}(t)=-\alpha\{1-\Delta t[{\rm Ci}(\Delta t)\sin(\Delta
t)-{\rm si}(\Delta t)\cos(\Delta t)]\}.\label{p8_7e8_24}
\end{equation}
This function, plotted along with the pole contribution in Fig.
\ref{fig1}b for the choice of parameters shown, has the following
features: $z_{bc1}(t)$ is negative for all $t$, it is
monotonically increasing, and its long-time behavior is
$z_{bc1}(t)\sim-2\alpha/(\Delta t)^2$. Also,
$z_{bc1}(t=0)=-\alpha$.

Let us survey, then, the peculiar features that this branch cut
contribution introduces into the time response $z(t)$. Visualizing
the $\epsilon=0$ spin-boson model as a symmetric double well
system coupled to its environment, the bc1 piece being negative
means that, if the system is initially in the left well, it will,
in the course of coherently tunnelling back and forth, spend more
time in the {\em right} well!  This effect becomes strongest at
long time, much longer than $T_2$, for in this regime the pole
contributions are exponentially small, while the bc1 contribution
decays like a power law. Experimentally it may be hard to see the
effect in this regime (on account of finite-temperature effects,
for example), so it is important to note that this memory effect
appears already at early times, indicating that already in the
first couple of coherent oscillations, there will be an excess
amplitude in the right-well excursions as compared with the
left-well excursions, by an amount proportional to $\alpha$. We
believe, on the basis of a variety of evidence\cite{foot2}, that
the Born approximation should be reliable up to $\alpha$'s of
order $1-2\%$; thus, experiments that look at coherent
oscillations accurately at the percent level (which, it seems,
will ultimately be necessary for performing quantum computation)
could readily see this bc1 effect.

We note several other interesting features of our solution for
$\epsilon=0$.  Taking into account the non-Markovian effects, we
can do a more precise calculation of the pole positions and
residues (only poles 3 and 4 contribute). We find, for $T=0$,
$\Gamma_2\equiv -{\rm Re}(s_3)=\Gamma_2^0 r$, where, as before
$\Gamma_2^0=\alpha \pi\Delta/2$, and the renormalization factor r
is given by $r =(1-\alpha)/(\kappa^2+\alpha^2\pi^2)<1$, with
$\kappa=1-2\alpha(1/2+C+\ln(\Delta/\omega_c))$. Further, ${\rm
Im}(s_3)=E+\delta E^{\rm Lamb}{\tilde r}$, with ${\tilde
r}=(\kappa-\alpha\pi^2/2(C+\ln(\Delta/\omega_c)))/(\kappa^2+(\alpha\pi)^2)$.
These expressions are obtained as systematic expansions in the
small parameters $\Gamma_2/E$ and $\delta E/E$, and they match a
direct numerical evaluation of the pole positions very well up to
$\alpha$'s of a few percent.
 For the corresponding pole residues we find the simple result in leading order
 $r_3+r_4=1+\alpha +O(\alpha^2)$.  This would be impossible in a Markovian
 theory, in which $z(t=0)=r_3+r_4$, so that $r_3+r_4$ would be
 exactly 1 to all orders in $\alpha$.  In fact this excess pole
 residue is exactly what is needed to cancel out the initial
 value of the bc1 contribution to $z(t)$.
We note that our results for the residues differ
from the weak-coupling expressions in the
literature\cite{weiss} (we are not aware of prior reports on $r,{\tilde
r}$).

For the biased model ($\epsilon\neq 0$) the bc2 contributions
become nonzero; we find that they give other peculiar
non-exponential corrections to the solution $z(t)$, very different
from the bc1 contribution.  We can do a nearly analytic evaluation
of the bc2 contribution to Eq. (\ref{DDV1_6_03}):  Using Eq.
(\ref{p2_11e2_543}) and expanding to lowest order in $\alpha$, we
find for the integrand of the sum of the $k=2$ and 3 terms of
(\ref{DDV1_6_03}),
\begin{equation}
z_{bc2}(s=i\omega)\approx{2\Delta^2\over
\omega}{b^-(\omega)\over(E^2-\omega^2+b^+(\omega))^2+b^-(\omega)^2}.\label{p10_3e10_8}
\end{equation}
Here $b(i\omega \pm \eta)\equiv b^+(\omega)\pm ib^-(\omega)$,
$b(s)\equiv\alpha(d(s)+n(s)(s^2+E^2)/\Delta)$, where $d(s)$ and
$n(s)$ are given by $N(s)=\Delta+\alpha n(s)$ (see Eq.
(\ref{p2_11e2_544})) and $D(s)=s^2+E^2+\alpha d(s)$ (see Eq.
(\ref{p2_11e2_545})).  Since $b^-(\omega)=0$ for $|\omega|\leq E$,
it is reasonable to expect that $b^-$ will grow linearly as one
passes onto the branch cut; and, in fact, we find from numerical
study that a good ansatz is $b^-(\omega)=(E-\omega){\tilde
b}^-(\omega)$, with ${\tilde b}^-(\omega)$ being a weakly varying,
real function of $\omega/E$. With this, for $\omega$ of order $E$,
Eq. (\ref{p10_3e10_8}) simplifies to
\begin{equation}
z_{bc2}(s=i\omega)\approx-{\Delta^2{\tilde b}^-(E)\over 2E^3}{1\over
\omega-E}.\label{p10_4e10_13}
\end{equation}
We find that (\ref{p10_4e10_13}) should be valid for
$\omega>E+b^+(E)/2E$.  Using (\ref{p10_4e10_13}) we can do the branch
cut integral, which gives (for $t\leq 1/(\alpha E x_0)$ --- see
Appendix for an alternative approach),
\begin{equation}
z_{bc2}(t)\approx\alpha x_1\log(x_0 \alpha
Et)\cos(Et+\phi).\label{DDVbc2note4}
\end{equation}
Here the dimensionless constants $x_0=|b^+(E)|/2\alpha E^2=
|\delta E|/\alpha E$ and $x_1=\Delta^2{\tilde b}^-(E)/2\alpha E^3$.
Since $b^\pm\propto\alpha$, these constants are independent of
$\alpha$.  The last expression for $x_0$ comes from an evaluation
of $b^+(E)$: it is directly related to the energy renormalization
in the Markov approximation, $b^+(E)=2E\delta E^{\rm Lamb}$.

\begin{figure}[htb]
\epsfxsize=15cm \epsffile{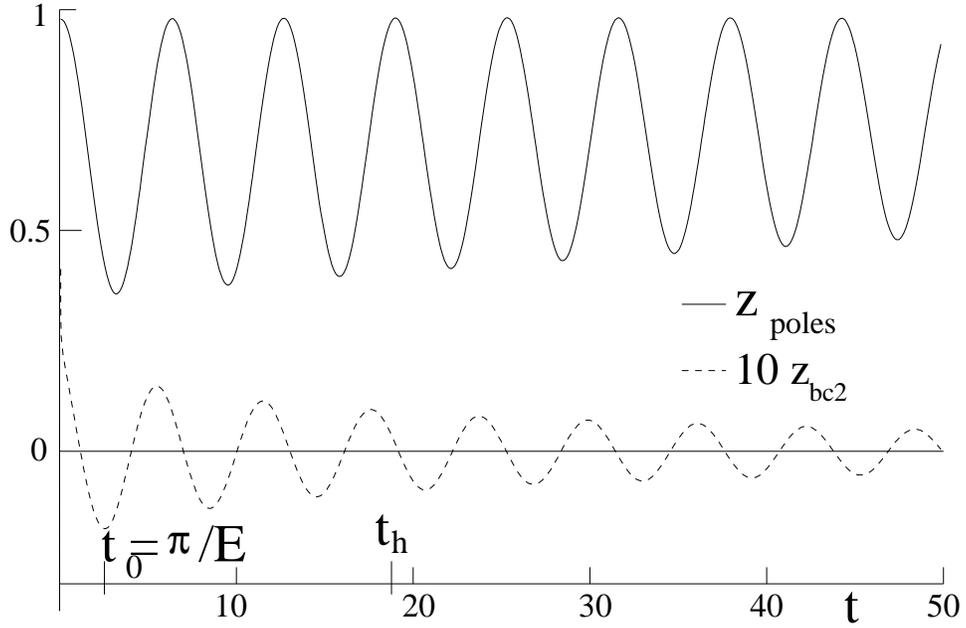} \caption{$z_{poles}(t)$ and
$z_{bc2}(t)$ for the biased case, illustrating the prompt loss of
coherence produced by bc2.  Here $E=1$,
$\epsilon/\Delta=-1.38$,
$\omega_c=30$,
$T=0$, and $\alpha=0.01$.  For these parameters, the time scale
for the prompt loss of coherence (using Eq.
(\protect\ref{p10_6e10_23})) is $t_h=18.98$.  $t_h$ is the time at
which the envelope of $z_{bc2}$ falls to half its value at
$t_0=\pi/E$.  This time scale is much shorter than the regular
exponential decay of coherence in $z_{poles}$; for our parameters,
$T_2=204.6$.} \label{fig2}
\end{figure}

In Fig.~\ref{fig2} we show a direct numerical evaluation of
$z_{bc2}(t)$. One can see the decay of the oscillatory part, which
is logarithmic according to Eq. (\ref{DDVbc2note4}).  Even though
the decay is very non-exponential, it is reasonable to attempt to
characterize this decay by a time scale.  Eq. (\ref{DDVbc2note4})
obviously does not work at $t=0$, since it is logarithmically
divergent. This is not surprising, since our calculation has
neglected cutoff effects (dependence on $\omega_c$), so Eq.
(\ref{DDVbc2note4}) is not expected to be correct for
$t<1/\omega_c$.  However, if we consider ``early" time to be the
first half-period of the coherent oscillation, $t_0=\pi/E$, then
Eq. (\ref{DDVbc2note4}) should be valid and we can use it to
characterize the decay by determining the time $t_h$ at which
$z_{bc2}(t)$ decreases to half its early-time value, i.e.,
$z_{bc2}(t_h)={1\over 2}z_{bc2}(t_0)$. We obtain
\begin{equation}
t_h={1\over  E}\sqrt{\pi E\over |\delta E|}\propto {1\over
E}{1\over\sqrt{\alpha}}\,.\label{p10_6e10_23}
\end{equation}
Surprisingly, $t_h\propto 1/\sqrt{\alpha}$ depends
non-analytically on $\alpha$. This explains the effect that is
evident in Fig. \ref{fig2}: for small $\alpha$, $t_h\ll T_2$, that
is, on the scale of $T_2$, there is a very rapid loss of coherence
as contributed by bc2.  This phenomenon may be called a {\em
prompt loss of coherence}, as it would appear experimentally as a
fast initial loss of coherence (from 100\% to $(1-c\alpha)100\%$,
$c$ being some constant near unity), followed by a much slower,
exponential decay of coherence on the regular $T_2$ time scale.

We give a few final remarks about the bc2 calculation.  The
absolute size of the bc2 contribution reaches a maximum near the
value of $\epsilon/\Delta$ used in Fig. \ref{fig2}; the relative
size of this contribution continues to increase as
$|\epsilon|/\Delta$ increases, so that it eventually becomes much
larger than the pole contribution (but all contributions to $z(t)$
go to zero as $|\epsilon|/\Delta\rightarrow\infty$).  When
$|\epsilon|\approx\Delta$, we find that, because of the prompt
loss of coherence, there is a {\em deficit} in the total pole
contribution, that is, $\sum_kr_k=1-O(\alpha)<1$.  Even in the
absence of an explicit branch cut computation, this deficit
signals the prompt loss of coherence, in that it indicates that
the exponentially decaying contributions to $z(t)$ do no account
for all the coherence near $t=0$.  Note that this is opposite to
the unbiased case, where, as a result of the bc1 part, there is an
{\em excess} pole contribution.

Naturally, many more regimes could be studied using the present
approach.  Recently, there has been interest in varying both the
system\cite{privman} and bath\cite{karl} initial conditions, as
well as in varying the model of the bath density of
states\cite{karl}.  For all these circumstances, the systematic
Born expansion procedure we report here can be done.  It is clear
on general grounds that the appearance of branch cut contributions
will not be restricted to the Ohmic model we have studied in
detail here; It is easy to show that for any spectral density of
the form $J(\omega)\propto\omega^n$ at low frequencies
($n=0,1,2,...$), $C(t)$ will have a power-law dependence at long
time, and thus $C(s)$ will have a branch point at $s=0$.  So,
interesting non-exponential contributions to the dynamics are
expected in all these cases.

Our hope is that, using the present and further exact calculations
of the weak-coupling behavior of the spin-boson model, a tool will
be made available to permit precision experiments to test the
validity of the model (which, at present, is only
phenomenologically justified) in various physical situations of
present interest in quantum information.  A fundamentally correct,
experimentally verified theory of the system and its environment
should ultimately be of great value in finding a satisfactory
qubit for the construction of a quantum information processor.

DPDV is supported in part by the National Security Agency and the
Advanced Research and Development Activity through Army Research
Office contract number DAAD19-01-C-0056.  He thanks the Institute
for Quantum Information at Cal Tech (supported by the National
Science Foundation under Grant. No. EIA-0086038) for its
hospitality during the initial stages of this work.  DL thanks the
Swiss NSF, NCCR Nanoscience, DARPA and the ARO.

\appendix
\section{Scaling form for branch cut integrals}

By numerical study we find that the branch cut integrals conform
to some simple scaling laws for small $\alpha$.  If we write the
bc1 and bc2 integrals as $z_{bc1}(t)=\int_0^\infty dx e^{-q_1xt}
z_{bc1}(s=q_1x)$ and $z_{bc2}(t)={\rm Re}\int_E^\infty dx e^{ixt}
z_{bc2}(s=ix)$, then we find that for small $\alpha$ and for
$s<<\omega_c$, $z_{bc1}(s)$ can be written in a scaling form
\begin{equation}
z_{bc1}(x)=(\alpha/E)f_1(\epsilon/\Delta,x/E).\label{DDVbc2note1}
\end{equation}
But for bc2 a very different scaling law applies:
\begin{equation}
z_{bc2}(x)=(1/E)f_2(\epsilon/\Delta,(x-E)/\alpha
E).\label{DDVbc2note2}
\end{equation}
Here $f_{1,2}$ are dimensionless, ``universal" functions that
govern the behavior of the branch cut contributions for small
$\alpha$.  For bc1, the behavior that the scaling law gives is
very simple: Eq. (\ref{DDVbc2note1}) implies that
$z_{bc1}(t)=\alpha {\bar f}_1(\epsilon/\Delta,Et)$, where ${\bar
f}_1$ is the Laplace transform of the scaling function $f_1$.  We
might have expected this behavior from Eq. (\ref{p8_7e8_24}), from
which we can read off the scaling function for $\epsilon=0$.  In
fact it appears from numerical studies that $f_1$ hardly changes
as $\epsilon$ is varied, except for an overall scale factor; that
is, ${\bar f}_1(\epsilon/\Delta,Et)\approx
a(\epsilon/\Delta)b(Et)$.  We find that the scaling function
$a(\tau)>0$ is peaked at $\tau=0$.  So, the memory effect
described above for $\epsilon=0$ persists for finite $\epsilon$,
but becomes smaller. For $|\epsilon|\approx\Delta$ the bc2
contribution, which we will describe now, becomes dominant over
the bc1 one.

Returning to Eq. (\ref{DDVbc2note2}), if we write the Fourier
transform of the scaling function as ${\bar
f}_2(\tau)=\int_0^\infty e^{ix\tau}f_2(x)dx$ and consider its
polar form ${\bar f}_2(\tau)=r_2(\tau)e^{i\phi_2(\tau)}$, then we
obtain
\begin{equation}
z_{bc2}(t)=\alpha r_2(\alpha Et)\cos(Et+\phi_2(\alpha
Et)).\label{DDVbc2note3}
\end{equation}
This shows that bc2 contributes an oscillatory part to the
solution, whose ``$T_2$" decay is determined by the features of
the scaling function $r_2$.  A few more observations about $f_2$
(obtained initially from numerical study) reveal some crucial
properties of the $r_2$ function: 1) $f_2(0)=0$;  2) $|f_2(x)|$
has a single maximum at $x=x_0$, where $x_0$ is some constant of
order unity; 3) Most important for the present discussion, for
$x>x_0$ $f_2(x)$ approaches $1/x$, that is, $f_2(x)\sim x_1/x$,
where $x_1$ is another real constant of order unity.  Fact 3)
implies that, for $\tau\rightarrow 0$, $r_2(\tau)\approx
x_1\log(x_0\tau)$.  That is, we conclude that at sufficiently
short time (actually for $t\leq 1/(\alpha E x_0)$, so a relatively
long time),
\begin{equation}
z_{bc2}(t)=\alpha x_1\log(x_0 \alpha
Et)\cos(Et+\phi),\label{DDVbc2note4app}
\end{equation}
as stated in the text.

\begin{thebibliography}{100}
\bibitem{foot1} Note that in this theory there is a very simple
relationship between $\sigma_y$ and $\sigma_z$:
$\Delta\sigma_y={\dot\sigma}_z$.
\end{thebibliography}
\end{document}